\documentclass[letterpaper]{article}
\usepackage{emulateapj}
\usepackage{apjfonts}

\begin{document}

\submitted{Low-resolution version.  High-resolution version and
machine-readable tables available upon request.  To appear in The
Astronomical Journal.}

\title{Stellar Cluster Fiducial Sequences with the Advanced 
Camera for Surveys$^1$}

\author{Thomas M. Brown, Henry C. Ferguson, Ed Smith}

\affil{STScI, 3700 San Martin Drive,
Baltimore, MD 21218;  tbrown@stsci.edu, ferguson@stsci.edu,
edsmith@stsci.edu} 

\author{Puragra Guhathakurta}

\affil{UCO / Lick Observatory, 271 Interdisciplinary Sciences
  Building, 1156 High Street, Santa Cruz, CA 95064; raja@ucolick.org}

\author{Randy A. Kimble, Allen V. Sweigart}

\affil{Code 667, NASA GSFC, Greenbelt, MD 20771; 
randy.a.kimble@nasa.gov, allen.v.sweigart@nasa.gov} 

\author{Alvio Renzini}

\affil{ESO, Karl-Schwarzschild-Strasse 2, 
Garching bei M$\ddot{\rm u}$nchen, Germany; arenzini@eso.org} 

\author{R. Michael Rich}

\affil{Division of Astronomy, Dpt.\ of Physics \& Astronomy, UCLA, Los
Angeles, CA 90095; rmr@astro.ucla.edu}

\author{Don A. VandenBerg}

\affil{Dpt.\ of Physics and Astronomy, University of Victoria, P.O. Box
3055, Victoria, BC, V8W 3P6, Canada; davb@uvvm.uvic.ca}

\begin{abstract}

We present color-magnitude diagrams of five Galactic globular clusters
and one Galactic open cluster spanning a wide range of metallicity
($-2.1 \lesssim $~[Fe/H]~$ \lesssim +0.3$), as
observed in the F606W (broad $V$) and F814W ($I$) bands with the
Advanced Camera for Surveys (ACS) on the {\it Hubble Space Telescope}.
These clusters are part of two large ACS programs measuring the star
formation history in the Andromeda halo, tidal stream, and outer disk.
In these programs, the clusters serve as empirical isochrones and as
calibrators for the transformation of theoretical isochrones to the
ACS bandpasses.  To make these data more accessible to the community,
for each cluster, we provide a ridge line tracing the stars on the main 
sequence, subgiant branch, and red giant branch, plus 
the locus of stars on the horizontal branch. In addition, we provide the
transformation of the Victoria-Regina isochrones to the ACS bandpasses.

\end{abstract}

\keywords{color-magnitude diagrams -- globular clusters: general --
globular clusters: individual (NGC~6341, NGC~6752, NGC~104, NGC~5927,
NGC~6528) -- open clusters: individual (NGC~6791)}

\section{Introduction}

The Advanced Camera for Surveys (ACS; Ford et al.\ 1998) on the {\it
Hubble Space Telescope (HST)} provides far greater broad-band optical
sensitivity than its predecessor, the Wide Field Planetary Camera 2
(WFPC2).  Two of the most widely used filters on the ACS Wide Field
Camera (WFC) are the F606W (broad $V$) and F814W ($I$); the former
provides three times the throughput of the analogous filter on
WFPC2, while the latter provides five times the throughput of the
corresponding filter on WFPC2.  Although the F555W on ACS more closely
approximates the Johnson $V$ bandpass, the F606W has far more grasp,
making it the filter of choice for deep imaging programs.

This enormous advance in sensitivity allows the {\it HST} to resolve
the old main sequence in populations out to the edge of the Local
Group, given a reasonable investment of exposure time ($\sim 100$
orbits) and a sufficiently sparse field ($\mu_V \gtrsim 26$~mag
arcsec$^{-2}$).  In order to measure the star formation history in the
Andromeda halo, we used this capability to obtain
extremely deep images of Andromeda in a
field 51$^\prime$ from the nucleus on the southeast minor axis (Brown
et al.\ 2003).  In addition to 
\linebreak

{\small \noindent $^1$Based on observations made with the NASA/ESA
Hubble Space Telescope, obtained at the Space Telescope 
Science Institute, 
which is operated by AURA, Inc., under NASA contract NAS 5-26555. These
observations are associated with proposals 9453 and 10265.}

\noindent
these deep images, our program included
brief exposures of five Galactic globular clusters, using the same
camera and filters: NGC~6341, NGC~6752, NGC~104, NGC~5927, and
NGC~6528.  These cluster observations 
serve two purposes: they provide empirical
isochrones for old simple stellar populations spanning a wide
range of metallicity, and they serve as calibrators for the
transformation of theoretical isochrones to the ACS bandpasses.
Subsequent to our first ACS program, we obtained additional
observations of the Andromeda tidal stream (discovered by Ibata et al.\
2001) and outer disk; this second program included an observation of
the old open cluster NGC~6791 in order to expand the range of
metallicities sampled in the empirical isochrones.  Because these data
would be useful references for other ACS programs, we tabulate here,
for each cluster, the ridge lines
tracing the main sequence (MS), subgiant branch (SGB), and red giant branch
(RGB) stars, plus the horizontal branch (HB) loci.
In addition, we describe our transformation of the Victoria-Regina
isochrones (hereafter VRI; Bergbusch \& VandenBerg 2001) to the ACS
bandpasses.  The VRI color-temperature relations have been revised
recently (VandenBerg \& Clem 2003), and the VRI grid has been extended
to higher metallicities and lower ages (VandenBerg, Bergbusch, \&
Dowler 2005).  The isochrone interpolation code and the models should
be available from the Canadian Astronomy Data Centre later this year.

\newpage

\hskip 0.8in
\parbox{4.0in}{
{\sc Table 1:}Parameters$^a$ of Galactic clusters observed with ACS

\begin{tabular}{lcclllr}
\tableline
                 & \multicolumn{2}{c}{exposures} &        &  &        &  \\
                 & F606W & F814W & $(m-M)_V$       & $E(B-V)$ &        & age \\
Name             & (sec) & (sec) & (mag)           &  (mag)   & [Fe/H] & (Gyr)\\
\tableline
\tableline
NGC~6341 (M92)   & 0.5079,5,90 & 0.5079,6,100 & 14.60 & 0.023& $-2.14$ & 14.5\\
NGC~6752         & 0.5079,4,40 & 0.5079,4,45  & 13.17 & 0.055& $-1.54$ & 14.5\\
NGC~104 (47~Tuc) & 0.5079,6,70 & 0.5079,5.5,72& 13.27 & 0.024& $-0.70$ & 12.5\\
NGC~5927         & 2,30,500 & 0.6934,15,340   & 15.85 & 0.42 & $-0.37$ & 12.5\\
NGC~6528         & 4,50,450 & 1,20,350        & 16.31 & 0.55 & $+0.00$ & 12.5\\
NGC~6791         & 0.5079,5,50 & 0.5079,5,50  & 13.50 & 0.14 & $+0.30$ &  9.0\\
\tableline
\end{tabular}

$^a$See \S4 for discussion and references for these parameters.
}

\vskip 0.2in

\section{Observations and Data Reduction}

The Galactic clusters in our programs are listed in Table~1.  The
globular clusters were observed as part of {\it HST} program GO-9453, while
NGC~6791 was observed as part of {\it HST} program GO-10265.  Although the
enormous sensitivity of ACS has enabled great strides in the deep imaging
of faint targets, ironically, the camera sensitivity makes it challenging to
observe relatively bright star clusters in our own Galaxy.  We
observed each cluster for one orbit, staggering the exposure times by
an order of magnitude to increase the dynamic range.  Due to the
exposure overheads (CCD readout, buffer dumps, etc.), only six ACS/WFC
images can be taken in a single orbit outside of the {\it HST} continuous
viewing zone, allowing three images in each bandpass (F606W and
F814W).  For the three relatively distant globular clusters (NGC~6341,
NGC~5927, and NGC~6528), we roughly centered the WFC on the cluster
core, maximizing the number of stars in the samples.  For the two
relatively nearby globular clusters (NGC~104 and NGC~6752), we offset
the ACS images from the cluster core, to include regions of lower
background and higher signal-to-noise ratio for the photometry of the
relatively faint main sequence stars.  The nearby open
cluster NGC~6791 is relatively sparse and subtends an area much larger
than the ACS field of view, so we centered the WFC roughly in the
cluster core.

We observed these clusters to create empirical isochrones and to
calibrate the transformation of theoretical isochrones to the ACS
bandpasses.  The observing strategy was designed to efficiently
achieve these purposes, but the data are not optimal for detailed
studies of the clusters themselves.  In particular, 
the exposures are brief (sometimes less than 1 sec), they are
not split into two subexposures, nor are they dithered, thus foregoing
the traditional method of cosmic ray rejection and precluding a better
sampled point spread function (PSF).  
However, by comparing each set of cluster images, even with
their different exposure times, we were able to create adequate
cosmic ray masks.  Note that, in 2003, the effective exposure times
for commanded exposures of less than 1 sec were remeasured and updated
in the {\it HST} calibration pipeline, such that a 0.5 sec exposure is really
0.5079 sec, and a 0.7 sec exposure is really 0.6934 sec (Gilliland \&
Hartig 2003).  Because the images were not dithered, 
no large changes in plate scale were applied when the
masked images were registered and coadded, 
as often done to better sample the PSF.

All of the images of a given cluster in a given bandpass were coadded
using the DRIZZLE package (Fruchter \& Hook 2002), with
masking of cosmic rays, saturated pixels, and bad pixels.  Although no
gross changes to the plate scale were applied, this step does correct
for the geometric distortion, small temporal changes in plate scale
due to velocity aberration and telescope breathing, and a small
difference in plate scale \linebreak

\vskip 1.75in

\noindent
between the two bandpasses.  Because the
images are filled with thousands of stars, we used the stellar
positions in each exposure to accurately determine the shifts and
scales needed to register the exposures.  The DRIZZLE package also
provides software for the masking of cosmic rays.  We tuned the
software parameters to aggressively mask cosmic rays without masking
the cores of unsaturated bright stars; these masks were then confirmed
by visual inspection.  In the medium and long exposures, the mask for
all saturated pixels (either due to a bright star or cosmic ray) was
enlarged with a border of 7 pixels to mask the bleeding of saturated
pixels into neighboring pixels; the short exposure suffers from very
few cosmic-ray hits and very little bleeding of saturated pixels.

The five globular clusters were all observed using a CCD gain of 1
e$^-$ per data number (DN), in order to match exactly the observing
mode used in our deep imaging of the Andromeda halo.  The observations were
planned before the installation of ACS, when it was unclear how much
uncertainty there would be in the relative gain corrections on the
camera.  Subsequent calibration programs have accurately measured the
gain correction, and showed that a gain of 2 e$^-$/DN offers significant
advantages when observing bright stars, with little increase in
quantization noise.  With a gain of 1 e$^-$/DN, the analog-to-digital
converter (ADC) saturates at 65,535 DN, prior to the CCD full well of
84,700 e$^-$, making it difficult to recover the flux from a saturated
star.  With a gain of 2 e$^-$/DN, fluxes can be measured not only to
the CCD full well, but beyond, via sampling the neighboring pixels
where charge ``bleeds'' from the saturated pixel (i.e., e$^-$ are
conserved in the conversion to DN). Because a gain of 2 e$^-$/DN can
significantly extend the dynamic range of the CCD, we observed
NGC~6791 with a gain of 2 e$^-$/DN.  As it turns out, the scarcity of giant
stars in our chosen field made this choice somewhat moot.

The undersampled images do not lend themselves to accurate
PSF-fitting; therefore, we obtained aperture photometry using the DAOPHOT
package (Stetson 1987).  We experimented with a range of aperture
sizes, and ultimately chose two different apertures for the
unsaturated and saturated stars.  For the unsaturated stars, we chose
a circular aperture of radius 2.5 pixels (0.125$\arcsec$) and a sky
annulus of radii 7--18 pixels.  With a gain of 1 e$^-$/DN, saturated
stars cannot be accurately photometered in a circular aperture because
charge is not conserved, so instead we measured the flux in a circular
annulus of radii 2.5--3.5 pixels, again with a 7--18 pixel sky
annulus.  We discarded from the catalog saturated stars that bled
beyond a 2.5 pixel radius, or if they were adversely
affected by bad pixels or were blended with bright neighbors.  The
photometry of the saturated stars was normalized to the same zeropoint
as the unsaturated stars by comparing photometry of the unsaturated
stars in the circular aperture and the annular aperture.  The
photometric catalog was then corrected to true apparent magnitudes
using TinyTim models of the {\it HST} PSF (Krist 1995) and observations of
the standard star EGGR 102 (a $V=12.8$~mag DA white dwarf) in the same
filters, with agreement at the 1\% level.

Charge transfer inefficiency (CTI) can be a problem for aging
large-format CCDs in the space radiation environment, 
causing stars to appear fainter than they actually
are.  The ACS detector consists of two chips, $4144 \times 2068$
pixels each, separated by a small horizontal gap.  Stars which fall
closer to the gap undergo more parallel transfers when the detector is
read, and thus suffer from more charge loss due to CTI.  The CTI
correction is approximately linear with the position of a star
relative to the gap, and approximately linear with the age of the
detector.  The correction is larger for faint stars and smaller when
there is a significant background.  Because the globular clusters were
all observed shortly after the ACS launch, the CTI correction for
isolated stars would be negligible ($\ll 0.01$~mag), even at the faint
end of the ridge lines we define for each cluster color-magnitude
diagram (CMD), as discussed below.
Because the globular cluster images are significantly crowded, the CTI
is, in practice, even smaller than it would be for isolated stars.  We
thus applied no CTI correction to our globular cluster photometry.
Unlike the globular clusters, 
NGC~6791 was observed 2.6 years after launch, and its ACS images are
relatively sparse; because the CTI
correction is small but not completely negligible ($\sim 0.01$~mag
near the faint limit at chip center), we applied a CTI correction to our
NGC~6791 photometry, using the algorithm of Riess \& Mack (2005).

Our photometry is in the STMAG system: $m = -2.5 \times $~log$_{10}
f_\lambda -21.1$~mag, where $f_\lambda = $ e$^- \times {\rm
PHOTFLAM/EXPTIME}$, EXPTIME is the exposure time, and PHOTFLAM is
$7.906 \times 10^{-20}$ erg s$^{-1}$ cm$^{-2}$ \AA$^{-1}$ / (e$^{-}$
s$^{-1}$) for the F606W filter and $7.072 \times 10^{-20}$ erg
s$^{-1}$ cm$^{-2}$ \AA$^{-1}$ / (e$^{-}$ s$^{-1}$) for the F814W
filter.  The STMAG system is a convenient system because it is
referenced to an unambiguous flat $f_\lambda$ spectrum; an object with
$f_\lambda = 3.63 \times 10^{-9}$ erg s$^{-1}$ cm$^{-2}$ \AA$^{-1}$
has a magnitude of 0 in every filter.  Another convenient and
unambiguous system that is widely used is the ABMAG system: $m=-2.5
\times $~log$_{10} f_\nu -48.6$~mag; it is referenced to a flat
$f_\nu$ spectrum, such that an object with $f_\nu = 3.63 \times
10^{-20}$ erg s$^{-1}$ cm$^{-2}$ Hz$^{-1}$ has a magnitude of 0 in
every filter.  It is thus trivial and unambiguous to convert any of
the data presented herein from STMAG to ABMAG: for F606W,
ABMAG~=~STMAG~$-0.169$~mag, and for F814W, ABMAG~=~STMAG~$-0.840$~mag.
Although our photometry could be transformed to ground magnitude 
systems (e.g., Johnson $V$) for comparison to theoretical isochrones
as well as other data in the literature, such
transformations always introduce significant systematic errors (see
Sirianni et al.\ 2005).  Instead of converting {\it HST} data to ground
bandpasses so that they can be compared to models in ground
bandpasses, it is preferable to produce models in one of the {\it HST}
instrument magnitude systems, in either STMAG or ABMAG.

The CMDs of each cluster are shown in Figures 1 through 6, along with
the ridge line spanning the MS, SGB,
and RGB.  Each ridge line was created by defining
regions along the MS-SGB-RGB locus, and taking the median color and
magnitude in each region.  The size of the region was varied along the
locus to allow clipping of outliers while including most of the stars
appropriate for defining the ridge line. Larger regions were defined
in parts of the CMD where the locus of stars is relatively linear,
where the photometric scatter is significant (at the faint end), and
where stars are scarce (at the bright end).  Smaller regions were
needed where the locus curves significantly (otherwise the ridge line
would smooth over these features), but fortunately this is also where
the photometric errors are relatively reasonable and the CMD is
relatively well-populated (between the turnoff and the base of the
RGB).  Near the tip of the RGB, where the CMD is very sparse, a mean
was used instead of a median if less than 5 stars fell in the region.
Figures 1--6 show, at representative locations,
horizontal and vertical bars spanning the sizes
of these regions used for defining the ridge lines.
We have also highlighted the HB locus in each
cluster; these stars were simply selected by eye from the
obvious overdensity of points in the vicinity of the HB.  The observed
fiducials (ridge line and HB locus) are difficult to determine
accurately in NGC~6528, because the cluster suffers from high,
spatially variable reddening (Heitsch \& Richtler 1999), and in
NGC~6791, due to the scarcity of stars.  A dotted line in each figure
indicates where stars can become saturated if they are well-centered
on an ACS pixel.  Note that when these ridge lines were presented
previously (Brown 2005; Brown et al.\ 2003, 2004), the
ridge lines were not shown above the saturation point, due to
uncertainties in the correction for saturated stars with a gain of 1
e$^-$/DN.  Although we have taken care to correct the saturated stars,
the correction is not as precise as the correction that can be done
with a gain of 2 e$^-$/DN, because one is effectively extrapolating
from the wings of the PSF.  The correction is also quite large at the
tip of the RGB; for example, at m$_{F814W} = 10.5$~mag in the NGC~104
CMD, a star clipped at the limit of the ADC can exhibit $\sim$25\%
less DN than the number of e$^-$ actually generated on the ACS detector.

Although we have shown the globular cluster fiducials in several
papers and conference proceedings, we have provided the tabular data
to ACS observers upon request only.  Publication of the tables was
delayed until we obtained our NGC~6791 observations so that
we could present a complete cluster dataset that extended to super-solar
metallicity.  It is worth noting that,
in the interim, a separate group published ridge lines for these
globular cluster data (Bedin et al.\ 2005).  Besides our
addition of NGC~6791, there are several
differences between their work and that presented here, 
related to the photometric methods, handling of
saturated stars, and reddening.  More significantly, Bedin et al.\
(2005) used a magnitude system referenced to a spectrum of Vega
that is older than the recent one of Bohlin \&
Gilliland (2004).  They also compared the globular cluster data to a
different isochrone set (Pietrinferni et al.\ 2004), 
and only did so for three of the globular
clusters (NGC~6341, NGC~6752, and NGC~104), which is understandable,
given the larger uncertainties associated with the clusters at higher
metallicity.  The work we present here will aid in the interpretation
of past and future papers in our study of the star formation history
in the Andromeda halo ({\it HST} program GO-9453), disk, and tidal stream
({\it HST} program GO-10265). It will also facilitate the analyses of other
groups working with CMDs derived from the most popular ACS bandpasses
(F606W and F814W).  These tables are given in the following sections,
where we also discuss how the ridge lines can be transformed to account for
different levels of reddening.

\section{Transformation of the Victoria-Regina Isochrones}

We briefly described our transformation of the VRI in Brown et al.\
(2004) but elaborate here.  As distributed, the VRI provide, for a
simple stellar population, the physical parameters (effective
temperature, surface gravity, luminosity, and mass) \linebreak

\epsfxsize=6.5in \epsfbox{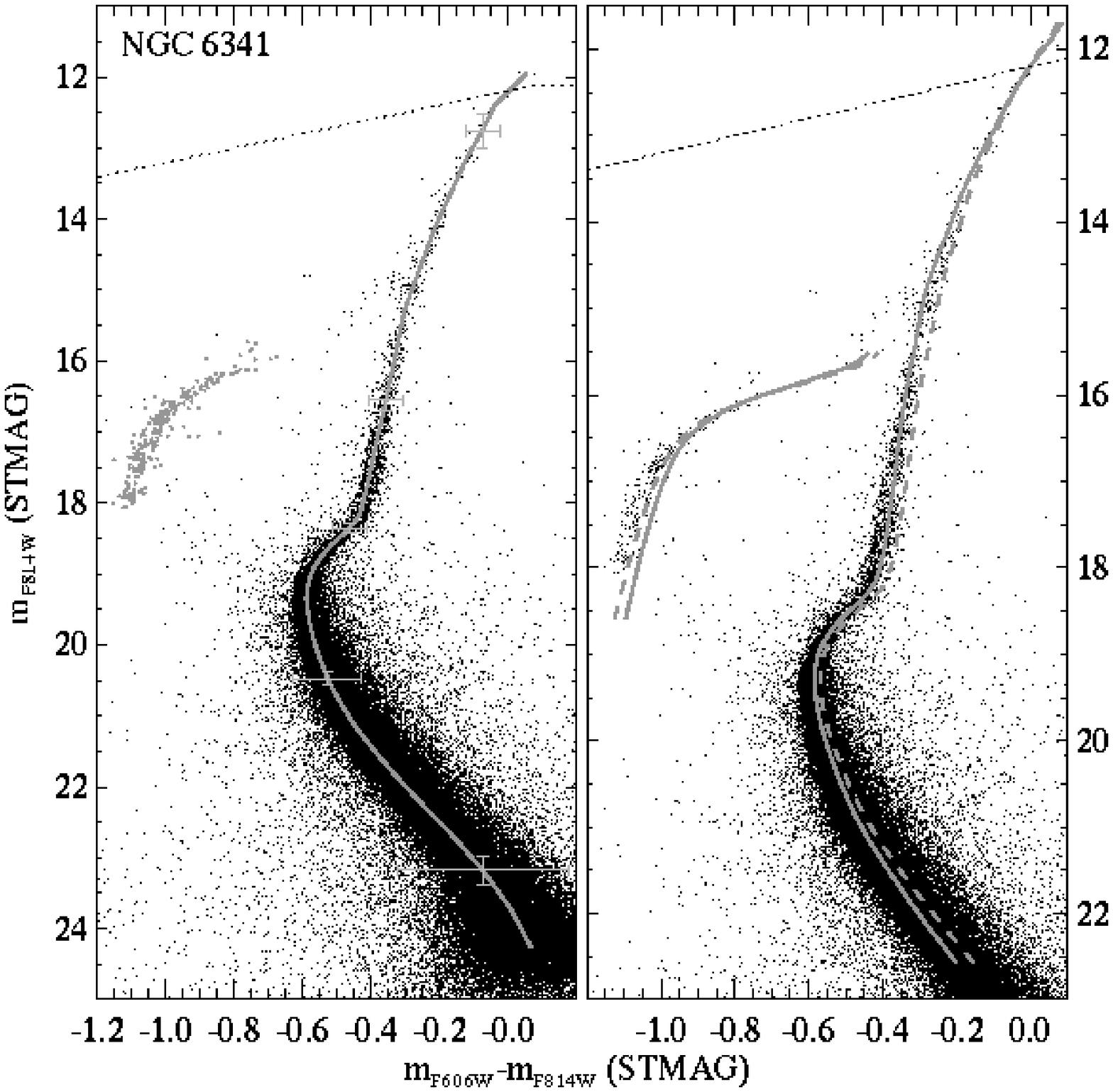}

\parbox{6.5in}{\small {\sc Fig.~1--}
{\it Left panel:} The CMD for NGC~6341, along with its ridge
line ({\it grey curve}).  The HB locus is highlighted in grey.  A
dashed line shows where stars can become saturated if well-centered
on an ACS pixel.  At representative points along the ridge line,
the size of the regions used to define the ridge line are shown
by horizontal and vertical lines.  {\it
Right panel:} The CMD for NGC~6341, along with an isochrone and ZAHB
sequence, shown with the empirical color correction ({\it solid
curve}) and without this correction ({\it dashed curve}).  Note that
the axes are not the same as those in the left panel, to better
show the level of agreement between the models and data.  The ZAHB
sequence shows better agreement with the data when no empirical color
correction is applied.}

\vskip 0.1in

\noindent
and the observed
magnitudes in ground bandpasses ($B$, $V$, $R$, and $I$).  
The adopted color-temperature relations for these ground bandpasses 
are the ones described by VandenBerg \& Clem (2003)$^2$\
and, as shown by VandenBerg (2000), they yield synthetic
CMDs that agree very well with the observed CMDs for a number of
well-studied Galactic clusters.
For this reason, Brown et al.\ (2003) used the synthetic
spectra of Lejeune, Cuisinier, \& Buser (1997) to calculate a
differential transformation between the ground bandpasses and the
corresponding ACS
bandpasses, $V - m_{F606W}$ and $I - m_{F814W}$, for all points on the
VRI, and then applied those differences to the isochrones, thus
producing a set of isochrones in the ACS bandpasses.  Comparison to
the ACS observations of Galactic \linebreak

\vskip 7.4in

\noindent
globular clusters showed that a small
empirical color correction ($\lesssim 0.05$~mag) was required to force
agreement between the transformed isochrones and the observations.  In
a subsequent analysis of the Andromeda globular cluster SKHB-312,
Brown et al.\ (2004) found that a direct transformation from physical
parameters to ACS bandpasses (instead of the 
\linebreak

{\small \noindent
$^2$These transformations are nearly identical with those reported by
Bell \& Gustafsson (1989) for turnoff stars.  At cooler temperatures,
redward adjustments to the synthetic colors were applied in order
that the predicted MS and RGB slopes agreed well with those observed.
The color-temperature relations derived from Kurucz model atmospheres
(e.g., see Castelli 1999) were used for stars hotter than $\approx
7000$ K.}

\epsfxsize=6.5in \epsfbox{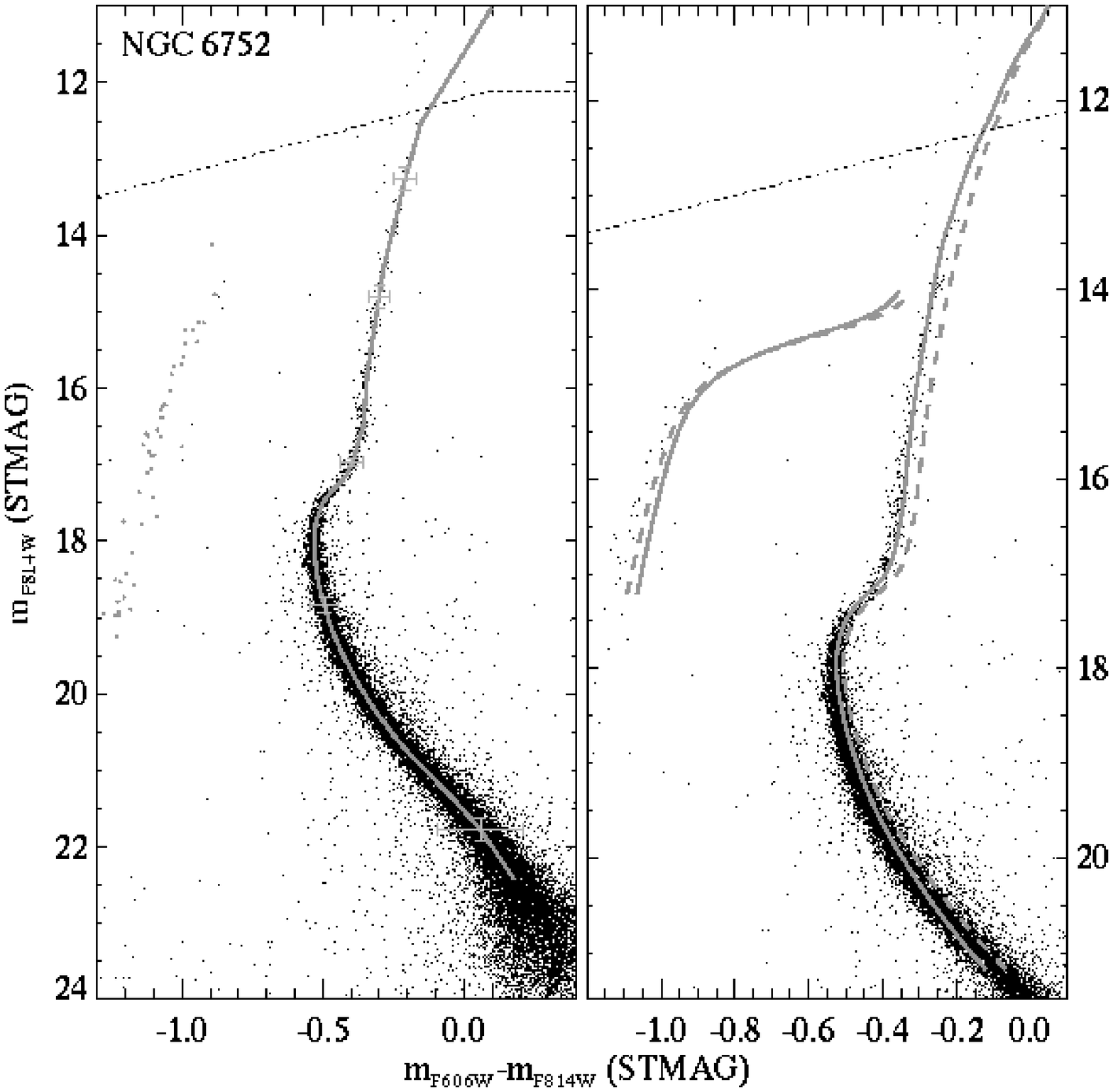}

\parbox{6.5in}{\small {\sc Fig.~2--} The same as Figure 1, but for NGC~6752.}

\vskip 0.1in

\noindent
differential transformation above) was slightly preferable, when done 
in conjunction with a more extensive grid of synthetic spectra (Castelli
\& Kurucz 2003).  The Castelli \& Kurucz (2003) grid provides spectra
over a wide range of metallicity, with and without alpha-enhancement,
so that the chemical compositions in the spectra can be well-matched
to those in the isochrones (Lejeune et al.\ 1997 provide only
scaled-solar models).  Isochrones transformed in this
new direct manner, when compared to the ACS Galactic cluster data,
still demonstrated the need for an empirical color correction, but it
was somewhat smaller and the functional form was simpler than that
required in Brown et al.\ (2003).  In the end, the result was very
similar to that obtained by Brown et al.\ (2003), because in both
cases the isochrones were forced to agree with the same set of
observational data.

To transform the isochrones to the ACS bandpasses,
we first interpolate the synthetic spectra grid of Castelli
\& Kurucz (2003) in metallicity, effective temperature, and surface gravity 
\linebreak

\vskip 6.63in

\noindent
to produce a spectrum at each point on the isochrone, and then
redden that spectrum using the curve of Fitzpatrick (1999).  Although
extinction is often handled in the literature by a ``reddening
vector'' that is constant over the full range of a CMD, in reality the
reddening produces a change in flux in each bandpass that depends upon
the spectral energy distribution of the star; thus, it is more
accurate to calculate the reddening at each point on the isochrone.
The reddened spectrum is then converted from an energy spectrum (erg
cm$^{-2}$ s$^{-1}$ \AA$^{-1}$) to a photon spectrum (photon cm$^{-2}$
s$^{-1}$ \AA$^{-1}$), multiplied by the throughput of each bandpass,
and integrated over wavelength, to produce the expected count rate on
the detector (e$^{-}$ s$^{-1}$).  Finally, this count rate is
converted to STMAG, using the same zeropoints given in \S2.  Because
the synthetic spectra grid of Castelli \& Kurucz (2003) does not
extend below 3500~K, the isochrone is truncated for any points below
this temperature (near the RGB tip).

Note that the VRI do not include He diffusion, which would 
\linebreak

\epsfxsize=6.5in \epsfbox{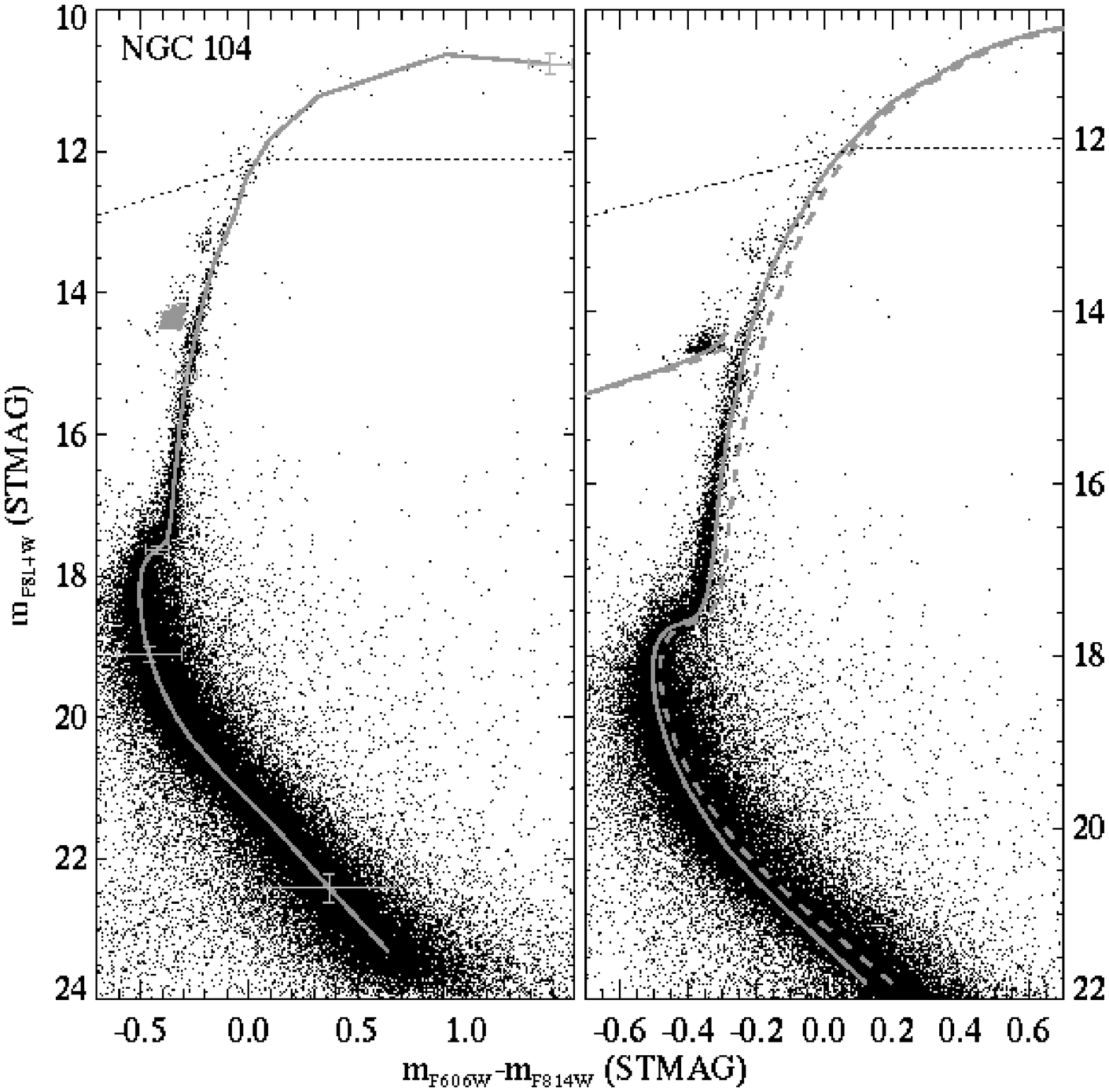}

\parbox{6.5in}{\small {\sc Fig.~3--} The same as Figure 1, but for NGC~104.}

\vskip 0.1in

\noindent
decrease their ages at a given turnoff luminosity by $\sim 10$\%,
thus avoiding discrepancies with the age of the Universe (VandenBerg
et al.\ 2002).  Although the ages of isochrones with He diffusion are
likely more accurate, if diffusion is allowed to act efficiently on other
elements in the surface layers,
such models show significant discrepancies when compared to observed CMDs.
For example, they 
fail to explain either the Li abundance versus effective temperature
relationship obeyed by field Population II dwarfs (see Richard et al.\
2002 and references therein) or the lack of any detectable difference
in the derived abundances between globular cluster stars 
at the turnoff and on the lower giant branch
(Gratton et al.\ 2001; James et al.\ 2004).  Apparently, there
must be some competing processes at work (e.g., turbulence
at the base of the convective envelope, as invoked by Richard et al.\ 2002)
that reduces the efficacy of diffusion in the surface layers of
metal-deficient stars.  However, the age effect is mainly due to the
settling of He in the stellar 
\linebreak

\vskip 6.75in

\noindent
core, and presumably this still occurs
at close to expected rates.  Although the best available models to use
in comparisons with stellar data are arguably those by Richard et al.\ (2002),
which take diffusion and turbulence into account, they have (so far)
been computed for only a few values of [Fe/H] and only as far as the lower
giant branch.  In view of these considerations, it seems advisable to
fit non-diffusive models to observed CMDs, and to reduce the ages so
obtained by $\approx 10$\% in order to provide the best estimates of
cluster ages.

\section{Cluster -- Isochrone Comparison}

The comparison of transformed isochrones to observed clusters is not
completely straightforward, because for even the most well-studied
clusters, there are still significant uncertainties in their parameters:
age, chemical composition, reddening, and distance.  We evaluated the
various values for these parameters in the literature, and settled
upon those (Table~1) 
\linebreak

\epsfxsize=6.5in \epsfbox{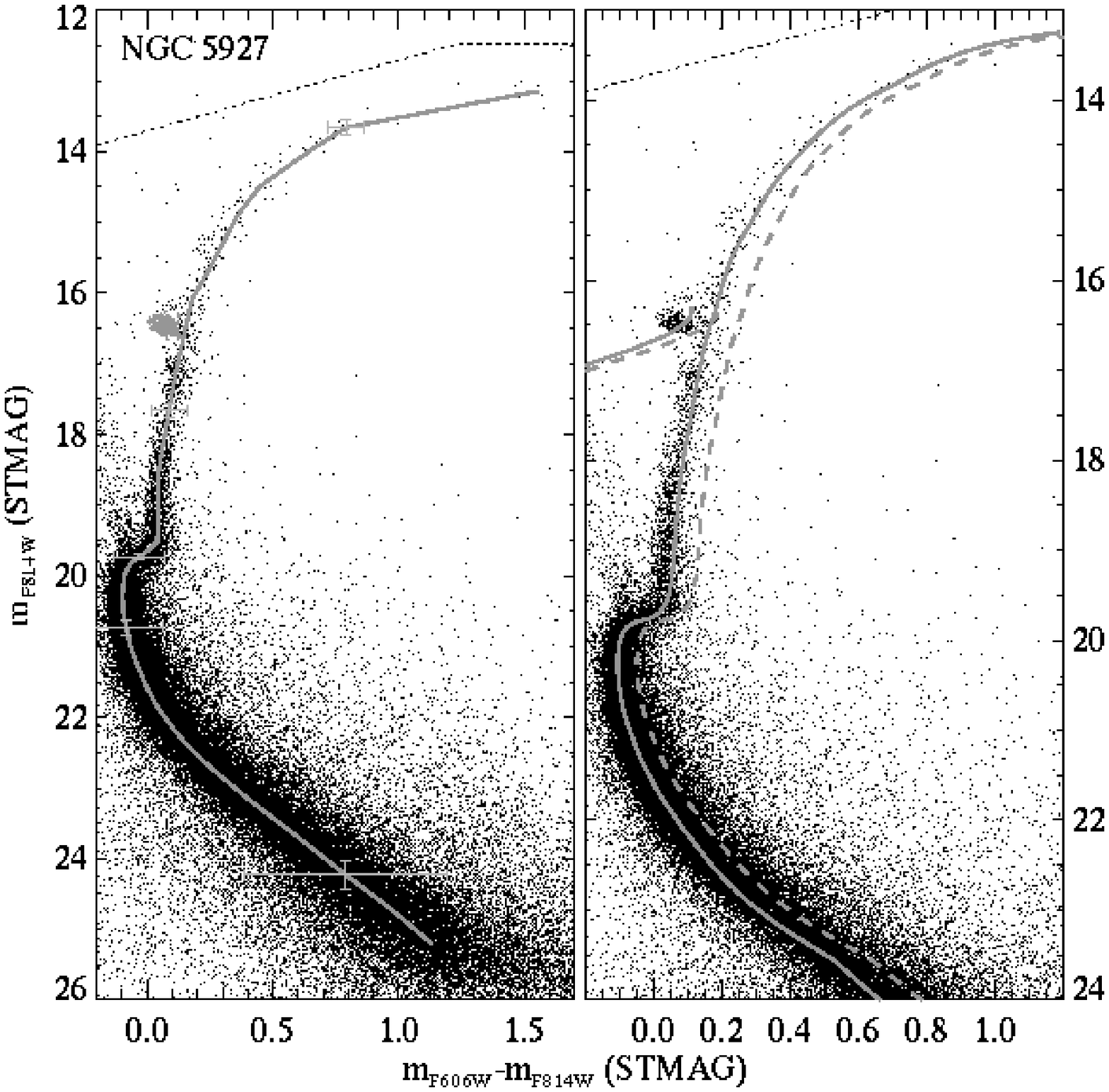}

\parbox{6.5in}{\small {\sc Fig.~4--} The same as Figure 1, but for NGC~5927.}

\vskip 0.1in

\noindent
that minimized the empirical correction of the
transformed isochrones and allowed a uniform correction for all
clusters.  If the same empirical correction can be applied to the
transformed isochrones associated with each of these clusters, then this
systematic offset can be reasonably attributed to systematic errors in
the isochrones, synthetic spectra, and calibration of the bandpasses.
In general, our values for the distance, [Fe/H], and reddening come
from the literature, while the age for each cluster is that which
provides the best agreement between the ACS data and the isochrone
in the vicinity of the turnoff.  As distributed, the VRI interpolation 
code produces isochrones over a continuous
range of age at discrete metallicities, but the sampling of the
metallicity grid is fine enough that isochrones can be matched to
clusters well within 0.1 dex.  For each of the globular clusters, we
compare to an isochrone with [$\alpha$/Fe]~=~+0.3 and a metallicity
closest to that in Table~1 (see Maraston et al.\ 2003 for
observational evidence of $\alpha$-enhancement in Galactic globular
\linebreak

\vskip 6.78in

\noindent
clusters over the full range of metallicity).  For the open cluster
NGC~6791, we interpolate the two isochrones nearest in metallicity
with [$\alpha$/Fe]~=~0.  Note that, to first order, isochrones without
alpha-enhancement look similar to isochrones with alpha-enhancement at
lower metallicity; this approximation is sometimes used in the
literature, although it is more accurate at lower metallicities than
at metallicities near the solar value.  Before comparing these isochrones
to the observed CMDs, we briefly discuss the parameters we adopted for 
each cluster.

{\bf NGC~6341.} For NGC~6341, we assumed the same distance, [Fe/H], and
reddening as used in Brown et al.\ (2003), but increased the age from
14~Gyr to 14.5~Gyr, which fits better under the new direct transformation
method.  VandenBerg \& Clem (2003) found good agreement between
the VRI and a $BV$ CMD of M92 (Stetson \& Harris 1988) when assuming
[Fe/H]~=~$-2.14$, $E(B-V) = 0.023$~mag, $(m-M)_V = 14.60$~mag, and an
age of 15~Gyr.  This value of [Fe/H] is very 
\linebreak

\epsfxsize=6.5in \epsfbox{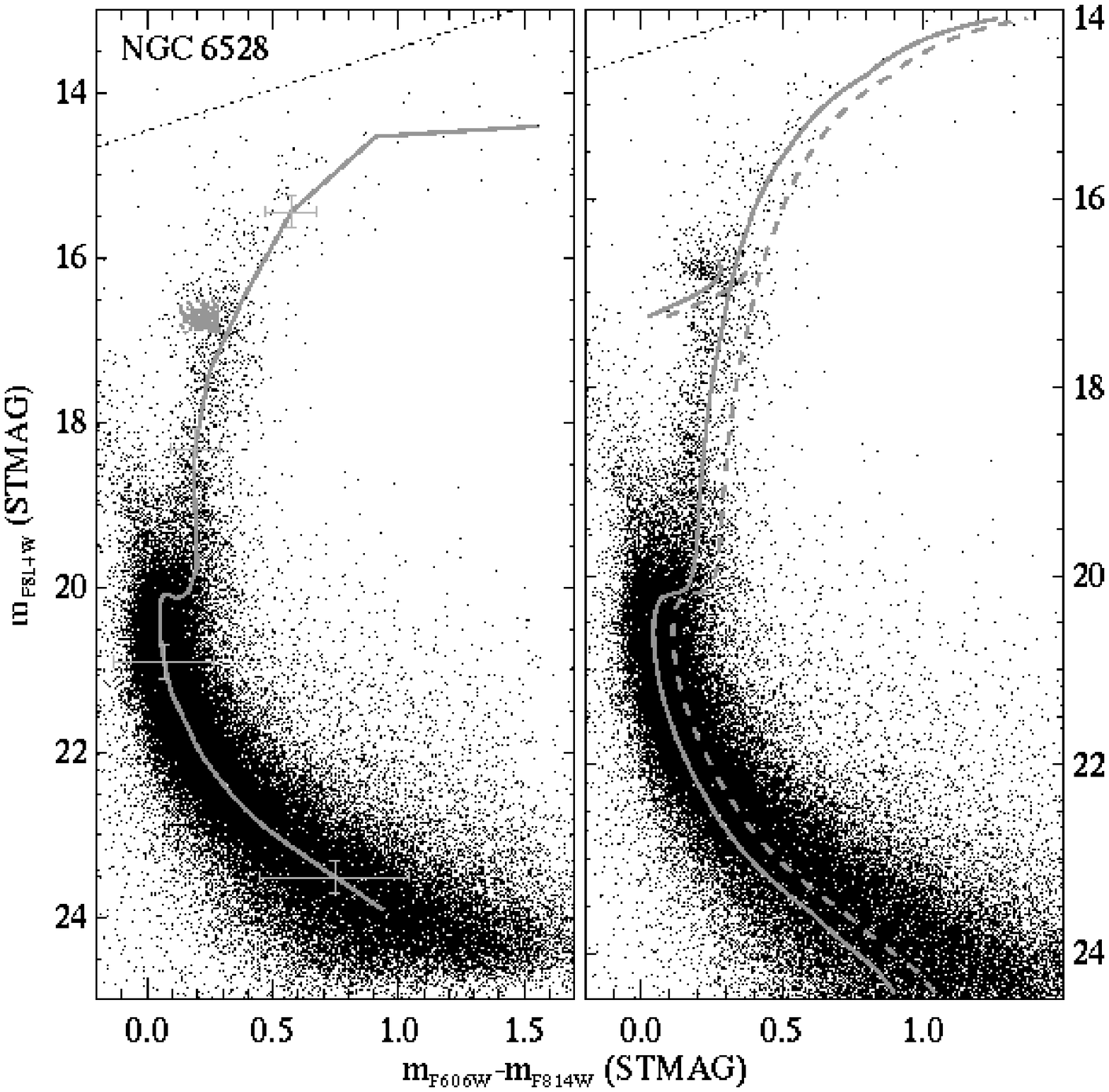}

\parbox{6.5in}{\small {\sc Fig.~5--} The same as Figure 1, but for NGC~6528.}

\vskip 0.1in

\noindent
close to that found
spectroscopically by Zinn \& West (1984; $-2.24 \pm 0.08$) and
Carretta \& Gratton (1997; $-2.16 \pm 0.02$).  For our comparison
between the ACS data and VRI (Figure 1), we used the
isochrones at [Fe/H]~=~$-2.14$.  The extinction comes from the dust
maps of Schlegel, Finkbeiner, \& Davis (1998), for a sight-line
through the Galaxy, which is reasonable for a distant halo cluster.
The distance comes from Grundahl et al.\ (2000), based upon the
metal-poor subgiant HD~140283; Grundahl et al.\ (2000) 
also found an age of 14.5~Gyr when
performing a distance-independent fit of the VRI 
to Str$\ddot{\rm o}$mgren photometry of NGC~6341.

{\bf NGC~6752.} For NGC~6752, we assumed the same distance, [Fe/H], and
reddening as used in Brown et al.\ (2003), but again increased
the age from 14~Gyr to 14.5~Gyr, which fits better under the new
direct transformation method.  VandenBerg (2000) found good
agreement between the VRI and the $BV$ CMD of
NGC~6752 (Penny \& Dickens 1986) when 
\linebreak

\vskip 6.77in

\noindent
assuming [Fe/H]~=~$-1.54$, which
is the same value found spectroscopically by Zinn \& West (1984;
$-1.54 \pm 0.09$). For our comparison between the ACS data and
VRI (Figure~2), we used the isochrones at [Fe/H]~=~$-1.54$.
Renzini et al.\ (1996) found a true distance modulus of $(m-M)_0 =
13.05$~mag by comparison of the cluster white dwarf sequence to local
white dwarfs with accurate parallaxes; they assumed an extinction of
$E(B-V) = 0.04$~mag (Penny \& Dickens 1986), and thus found an
apparent distance modulus of $(m-M)_V = 13.17$.  We have updated the
extinction to $E(B-V)=0.055$~mag, using the dust maps of Schlegel et
al.\ (1998).  Our adopted age falls in the range found in the recent
literature; Renzini et al.\ (1996) found an age of 15.5~Gyr using 
non-diffusive models and 14.5~Gyr using diffusive models, while
VandenBerg (2000) found an age of 12.5~Gyr, on the assumption of a 
somewhat larger distance modulus, using non-diffusive models.

{\bf NGC~104.} For NGC~104, we assumed the same parameters 
\linebreak

\epsfxsize=6.5in \epsfbox{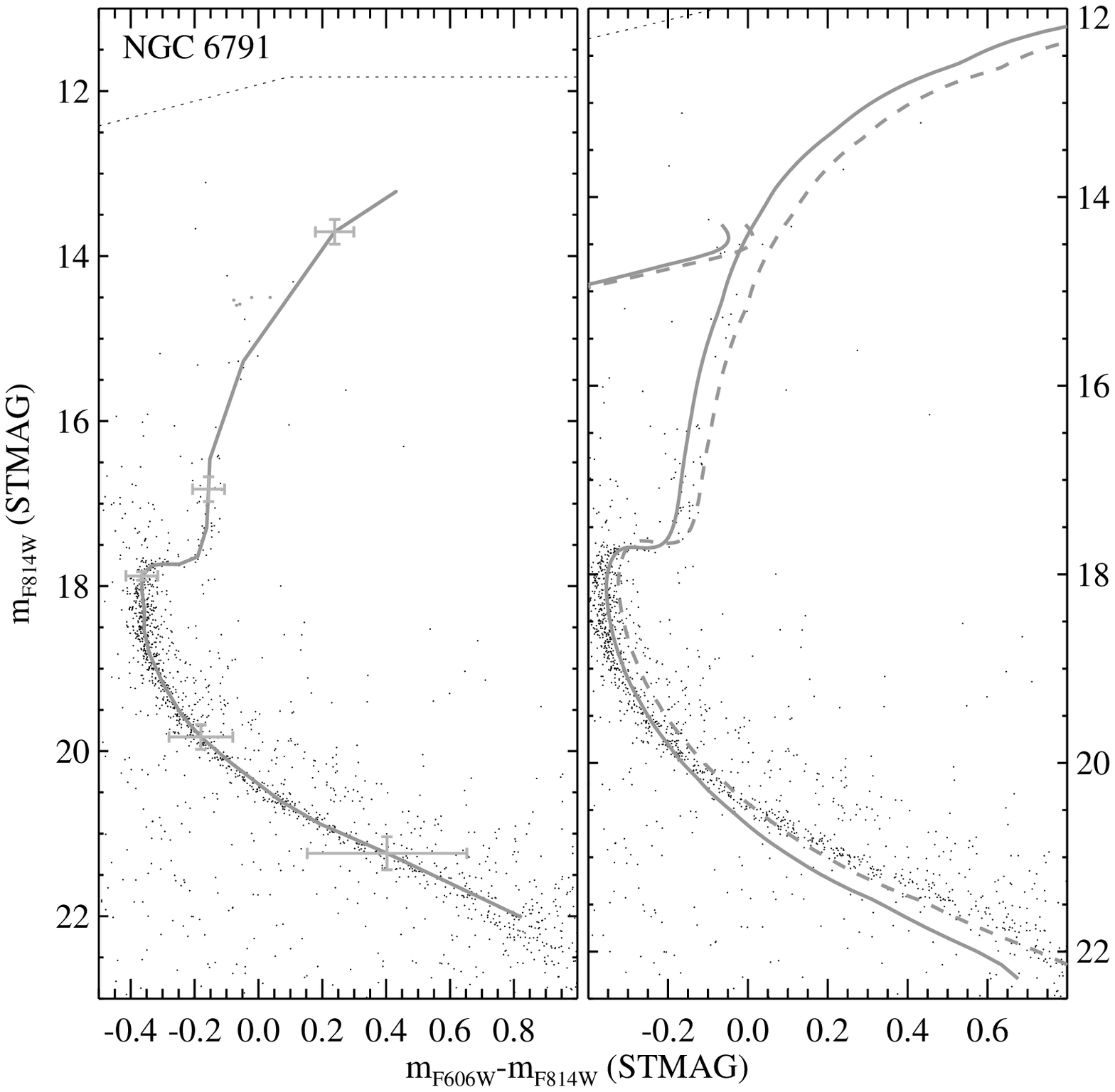}

\parbox{6.5in}{\small {\sc Fig.~6--} The same as Figure 1, but for NGC~6791.}

\vskip 0.1in

\noindent
as those
used in Brown et al.\ (2004).  The metallicity comes from the
high-resolution spectroscopy of Carretta \& Gratton (1997; $-0.70 \pm
0.03$), which agrees well with the [Fe/H]$_{\rm II}$ fits of Kraft \&
Ivans (2003; $-0.70 \pm 0.09$) and with the spectroscopic fits of Zinn
\& West (1984; $-0.71 \pm 0.08$). For our comparison between the ACS
data and VRI (Figure 3), we used isochrones at [Fe/H]~=~$-0.705$.
From fits to the cluster white dwarf sequence, Zoccali et al.\ (2001)
found an apparent distance modulus of $(m-M)_V = 13.27 \pm 0.14$~mag.
Gratton et al.\ (2003) present several different determinations for
$E(B-V)$, with an average value of 0.024~mag.  Our adopted age of
12.5~Gyr is midway between that found by VandenBerg \& Clem (2003;
12~Gyr), who adopted a slightly larger distance and smaller metal
abundance, and Zoccali et al.\ (2001; 12.9~Gyr with diffusive models and
13.5~Gyr with non-diffusive models).

{\bf NGC~5927.} For NGC~5927, we assumed the same parameters as those
used in Brown et al.\ (2004).  The metallicity is 
\linebreak

\vskip 6.77in

\noindent
that adopted by
Harris (1996), and represents a combination of various spectroscopic
measurements in the literature (Zinn 1985; Armandroff \& Zinn 1988;
Fran\c{c}ois 1991).  For our comparison between the ACS data and VRI
(Figure 4), we used isochrones at [Fe/H]~=~$-0.397$.
Brown et al.\ (2004) adopted values for distance
and reddening that were slightly different from those of Harris (1996),
who gives $(m-M)_V = 15.81$~mag and $E(B-V)=0.45$~mag, but the Brown
et al.\ (2004) values (15.85~mag and 0.42~mag, respectively) are well
within the uncertainties, given the high, spatially-variable reddening
(Heitsch \& Richtler 1999).  Our adopted age of 12.5~Gyr is also well
within the wide range of ages in the literature for this cluster
(see Feltzing \& Gilmore 2000 for a summary of recent estimates
for the age, metallicity, distance, and reddening).

{\bf NGC~6528.} For NGC~6528, we assumed the same distance, reddening,
and age as used in Brown et al.\ (2003), but we assumed a slightly
higher metallicity, [Fe/H]~=~0.0, which is still well within the range
of spectroscopic metallicities in the literature (Momany et al.\
2003).  For our comparison between the ACS data and VRI (Figure 5), we
used isochrones at [Fe/H]~=~0.  
As with the case of NGC~5927, NGC~6528 suffers from high,
spatially-variable reddening (Heitsch \& Richtler 1999 and references
therein), resulting in considerable uncertainties in its fundamental
parameters (see Feltzing \& Gilmore 2000).  Brown et al.\ (2003) found
that if NGC~6528 is assumed to lie at an apparent distance modulus of
$(m-M)_V = 16.15$~mag (Momany et al.\ 2003), the cluster, when shifted
to the Andromeda distance and reddening, is too faint by 0.16~mag,
implying $(m-M)_V = 16.31$~mag, which is the distance we have assumed
here.  We assumed the same extinction, $E(B-V) = 0.55$~mag, as found by
Momany et al.\ (2003) in their analysis of optical and near-infrared
imaging.  Our adopted age of 12.5~Gyr is nearly the same as that found
by Momany et al.\ (2003; 12.6~Gyr).  Because the metallicity and reddening
are very uncertain for NGC~6528, it is difficult to evaluate the 
empirical corrections needed to achieve agreement between the
isochrone and observed CMD, but those corrections are the same as
those applied to the other clusters.  

{\bf NGC~6791.} For NGC~6791, the Carney et al.\ (2005) analysis of
their infrared HB photometry implies $(m-M)_V =13.50$~mag
and $E(B-V) = 0.14$~mag, and we have assumed those values here.  The
metallicity for NGC~6791 is quite uncertain (see Taylor 2001 for an
extensive review), but Carney et al.\ (2005) find acceptable fits
to their infrared photometry if they assume [Fe/H]~=~+0.5 (and an
age of 7.5 Gyr) or [Fe/H]~=~+0.3 (and and age of 9~Gyr); we find good
agreement with the latter pair of parameters.  This metallicity falls
midway between two metallicities in the VRI, so we interpolated between
isochrones at [Fe/H]~=~+0.23 and [Fe/H]~=+.37; the comparison between
the isochrone and ACS data is shown in Figure 6.

The comparisons between the data and isochrones are shown in Figures
1--6 ({\it right-hand panels}).  The dashed line shows the transformed
isochrone when the above steps are performed, but no empirical
correction is applied.  The isochrones are generally too red over much
of the CMD, except for the bottom of the main sequence.  Furthermore,
the color difference between the turnoff and the base of the RGB is
larger than observed, which may indicate, e.g., a preference 
either for models 
that treat diffusive processes and turbulence, or for models with higher
oxygen abundances, as both possibilities are ways of reducing the turnoff
temperatures without affecting the location of the RGB significantly.
While adopting older isochrones could also alleviate
this problem, the required ages are implausibly large. Because an accurate
determination of the age and metallicity in a CMD generally comes from
a fit of the upper MS, turnoff, SGB, and RGB, we apply a small
empirical color correction, specified as a first-order polynomial: \\
\begin{center}
$(m_{F606W}-m_{F814W})_{new} = 0.92 \times 
(m_{F606W}-m_{F814W})_{old} - 0.06.$\\
\end{center}
This correction gives good
agreement from the upper MS to a point on the RGB slightly above the HB,
but then the upper RGB increasingly deviates to the blue.  Thus, we
change the correction for stars at log~$g < 2$ to:\\
\begin{center}
$(m_{F606W}-m_{F814W})_{new} = 0.92 \times
(m_{F606W}-m_{F814W})_{old} - 0.06 + 0.05 \times (2-$ log~$g)$.
\end{center}
This correction gives good agreement to the end of the isochrone
(which, as stated above, can end before the RGB tip, due to the 3500~K
limit in the grid of synthetic spectra).  In general, the transformed
isochrones with the empirical correction (shown as a solid curve in
the right-hand panels of Figures 1--6) agree at the $\lesssim
0.02$~mag level with the observed ridge lines from a point 1.5~mag
below the MS turnoff through the upper RGB, which thus allows the
isochrones to be used for fitting the age and metallicity in ACS CMDs.
Over this range, the empirical correction is small ($\lesssim
0.05$~mag) but not insignificant.  In the two most metal-rich clusters
(NGC~6528 and NGC~6791), the agreement on the upper RGB might not be
as good as that in the more metal-poor clusters, but this is difficult
to evaluate, given the scarcity of stars in NGC~6791 and the high
differential reddening in NGC~6528.  We make no attempt to improve the
agreement between the isochrones and data on the lower MS, because
there is too much variation in the disagreement, from cluster to
cluster, to suggest a suitable correction.  Furthermore, photometry on
the lower MS is not required to measure ages in CMDs (and it is beyond
the reach of current instrumentation for stellar populations beyond
our Galaxy and its satellites).

We also show models of the zero-age HB (ZAHB) for each cluster, again
transformed using the method above.  For consistency with the isochrones,
the ZAHB loci were taken from
VandenBerg et al.\ (2000); their $M_V$ values in the center of the
instability strip agree well with those derived from studies of RR
Lyrae stars in Galactic globular clusters spanning a wide range in
metallicity (see De Santis \& Cassisi 1999; Cacciari et al.\ 2005).
The ZAHB should trace the bottom of the HB locus, because luminosity
increases during HB evolution.  In general, the ZAHB models show
better agreement with the data when no color correction is applied
({\it dashed}) than when the color correction is applied ({\it
solid}).  Although we do not use HB models in our fits of the
Andromeda star formation history, it appears that a transformation of
HB models to the ACS bandpasses should not employ the empirical color
correction we employ in the transformation of the isochrones.  Because
the uncorrected ZAHB agrees well across the entire range of color,
especially at the blue end, this might be indicating that a
significant part of the correction required for the isochrones
is driven by limitations in the
synthetic spectra of stars at low effective temperature and low
surface gravity, where one must account for large atomic and molecular
opacities.

The transformed models for each cluster can also be used to estimate
the effects of reddening on the observed fiducials, allowing them to
be transformed to any given distance and reddening, for comparison to
other CMDs.  For example, in Brown et al.\ (2004), we transformed the
fiducials of NGC~5927 and NGC~104 to the Andromeda distance and
reddening.  At each point on the fiducial, we took the closest point
on the associated model, calculated the position in the CMD at both
the cluster reddening and the Andromeda reddening, applied this
difference to the fiducial, and then shifted by the difference in
distance modulus.  The absorption in each of the ACS bandpasses
($A_{F606W}$ and $A_{F814W}$) is a function of both the effective
temperature of the source and the foreground reddening, as
parameterized by $E(B-V)$.  The ratio of absorption in an ACS band to
$E(B-V)$ is not constant: $R_{F606W} \equiv A_{F606W} / E(B-V)$ and
$R_{F814W} \equiv A_{F814W} / E(B-V)$ are both very slowly varying
functions of $E(B-V)$.  For example, if one assumed that $A_{F606W}$
at the NGC~104 turnoff was ten times larger for $E(B-V) = 1.0$~mag
than it was at $E(B-V) = 0.1$~mag, this would introduce an error of
0.08~mag.  Thus, the absorption in {\it HST} bandpasses has sometimes been
published in tables giving absorption in the bandpasses at specific
values of effective temperature and $E(B-V)$ (e.g., see Bedin et al.\
2005 for absorption in ACS bands and Holtzman et al.\ 1995 for
absorption in WFPC2 bands, both specified at two distinct effective
temperatures).  However, using the isochrone associated with each
ridge line, one can parameterize the absorption along the ridge line
more generally so that it can be transformed to any reference frame.
The parameterization is straightforward because $R_{F606W}$ and
$R_{F814W}$ are nearly linear functions of $E(B-V)$, so we provide, in
Tables 2--7, the ridge line for each cluster (as observed) plus the
coefficients ($\alpha$, $\beta$) needed to calculate $R_{F606W}$ and
$R_{F814W}$ along the ridge line:\\
\begin{center}
$R_{F606W} = \alpha_{F606W} + \beta_{F606W} \times E(B-V)$\\
$R_{F814W} = \alpha_{F814W} + \beta_{F814W} \times E(B-V)$.  
\end{center}
\noindent The absorption is then the product of
$R$ and $E(B-V)$.  We derive these coefficients by transforming the
corresponding isochrone to different values of $E(B-V)$, fitting
$R_{F606W}$ and $R_{F814W}$ as functions of $E(B-V)$, and then
matching points on the isochrone to points on the ridge line (with the
isochrone at the cluster extinction).  Where the ridge lines extend
beyond the isochrone at the faint and bright ends of the ridge line,
we simply take the closest isochrone point in the CMD.  Because
$R_{F606W}$ and $R_{F814W}$ are well-approximated by the coefficients
in Table 2--7, there is little discernible difference ($\lesssim
0.01$~mag) between a transformation of the ridge line using these
coefficients and the more exact method of Brown et al.\ (2004).  In
Tables 8--13, we provide the analogous data for the HB loci, again by
calculating the ZAHB model at different values of $E(B-V)$, fitting
the dependence of $R_{F606W}$ and $R_{F814W}$ on $E(B-V)$, and
matching stars in the HB locus to points in the ZAHB model at the same
color.  We stress that the ridge lines and HB loci in Tables 2--13
present the fiducials as observed, and thus they include the reddening
for each cluster (Table 1); to transform these fiducials to a different
reference frame with a distinct reddening, one must first subtract
the reddening intrinsic to each cluster and then add the reddening
for the new reference frame, using the coefficients in Tables 2--13.

\vskip 0.3in

\noindent
\parbox{3.25in}{
{\sc Table 2:}NGC~6341 ridge line and reddening coefficients

\begin{tabular}{cccccc}
\tableline
\tableline
$m_{F606W}$ & $\alpha_{F606W}$ &  & $m_{F814W}$ & $\alpha_{F814W}$ &  \\ 
(mag)       &  (mag)           & $\beta_{F606W}$ & (mag) & (mag) & $\beta_{F814W}$  \\
\tableline
24.346 &  2.628 &  -0.077 & 24.279 &  1.709 &  -0.025 \\
23.731 &  2.628 &  -0.077 & 23.724 &  1.709 &  -0.025 \\
23.103 &  2.628 &  -0.077 & 23.178 &  1.709 &  -0.025 \\
22.570 &  2.628 &  -0.077 & 22.726 &  1.709 &  -0.025 \\
22.090 &  2.639 &  -0.078 & 22.327 &  1.712 &  -0.025 \\
21.592 &  2.654 &  -0.080 & 21.909 &  1.715 &  -0.025 \\
21.194 &  2.666 &  -0.082 & 21.572 &  1.718 &  -0.025 \\
20.712 &  2.679 &  -0.083 & 21.159 &  1.721 &  -0.025 \\
20.248 &  2.689 &  -0.084 & 20.747 &  1.723 &  -0.025 \\
19.947 &  2.695 &  -0.084 & 20.475 &  1.725 &  -0.025 \\
\tableline
\end{tabular}
Note: Table 2 is presented in its entirety in the electronic edition
of the Astronomical Journal.  A portion is shown here for guidance
regarding its form and content.
}\\ \\ \\ \\ \\ \\ \\ 

\noindent
\parbox{3.25in}{
{\sc Table 3:}NGC~6752 ridge line and reddening coefficients

\begin{tabular}{cccccc}
\tableline
$m_{F606W}$ & $\alpha_{F606W}$ &  & $m_{F814W}$ & $\alpha_{F814W}$ &  \\ 
(mag)       &  (mag)           & $\beta_{F606W}$ & (mag) & (mag) & $\beta_{F814W}$  \\
\tableline
22.610 &  2.613 &  -0.074 & 22.432 &  1.707 &  -0.025 \\
22.178 &  2.613 &  -0.074 & 22.066 &  1.707 &  -0.025 \\
21.831 &  2.613 &  -0.074 & 21.773 &  1.707 &  -0.025 \\
21.398 &  2.613 &  -0.074 & 21.424 &  1.707 &  -0.025 \\
20.970 &  2.620 &  -0.075 & 21.084 &  1.708 &  -0.025 \\
20.609 &  2.633 &  -0.077 & 20.799 &  1.711 &  -0.025 \\
20.295 &  2.644 &  -0.079 & 20.544 &  1.713 &  -0.025 \\
19.874 &  2.658 &  -0.080 & 20.194 &  1.716 &  -0.025 \\
19.526 &  2.669 &  -0.082 & 19.897 &  1.719 &  -0.025 \\
19.169 &  2.679 &  -0.083 & 19.585 &  1.721 &  -0.025 \\
\tableline
\end{tabular}
Note: Table 3 is presented in its entirety in the electronic edition
of the Astronomical Journal.  A portion is shown here for guidance
regarding its form and content.  
}

\vskip 0.63in

\noindent
\parbox{3.25in}{
{\sc Table 4:}NGC~104 ridge line and reddening coefficients

\begin{tabular}{cccccc}
\tableline
$m_{F606W}$ & $\alpha_{F606W}$ &  & $m_{F814W}$ & $\alpha_{F814W}$ &  \\ 
(mag)       &  (mag)           & $\beta_{F606W}$ & (mag) & (mag) & $\beta_{F814W}$  \\
\tableline
\tableline
23.966 &  2.567 &  -0.067 & 23.323 &  1.692 &  -0.024 \\
23.332 &  2.567 &  -0.067 & 22.836 &  1.692 &  -0.024 \\
22.784 &  2.567 &  -0.067 & 22.415 &  1.692 &  -0.024 \\
22.077 &  2.567 &  -0.067 & 21.867 &  1.692 &  -0.024 \\
21.519 &  2.578 &  -0.068 & 21.436 &  1.698 &  -0.025 \\
21.028 &  2.594 &  -0.071 & 21.067 &  1.704 &  -0.025 \\
20.466 &  2.618 &  -0.075 & 20.641 &  1.710 &  -0.025 \\
20.022 &  2.636 &  -0.077 & 20.293 &  1.713 &  -0.025 \\
19.771 &  2.647 &  -0.079 & 20.088 &  1.715 &  -0.025 \\
19.467 &  2.658 &  -0.080 & 19.832 &  1.717 &  -0.025 \\
\tableline
\end{tabular}

Note: Table 4 is presented in its entirety in the electronic edition
of the Astronomical Journal.  A portion is shown here for guidance
regarding its form and content.  
}

\vskip 0.63in

\noindent
\parbox{3.25in}{
{\sc Table 5:}NGC~5927 ridge line and reddening coefficients

\begin{tabular}{cccccc}
\tableline
$m_{F606W}$ & $\alpha_{F606W}$ &  & $m_{F814W}$ & $\alpha_{F814W}$ &  \\ 
(mag)       &  (mag)           & $\beta_{F606W}$ & (mag) & (mag) & $\beta_{F814W}$  \\
\tableline
\tableline
26.339 &  2.560 &  -0.066 & 25.207 &  1.679 &  -0.024 \\
25.678 &  2.560 &  -0.066 & 24.710 &  1.679 &  -0.024 \\
25.018 &  2.560 &  -0.066 & 24.229 &  1.679 &  -0.024 \\
24.284 &  2.564 &  -0.067 & 23.688 &  1.685 &  -0.024 \\
23.600 &  2.579 &  -0.068 & 23.188 &  1.696 &  -0.024 \\
22.972 &  2.601 &  -0.072 & 22.719 &  1.706 &  -0.025 \\
22.554 &  2.618 &  -0.075 & 22.396 &  1.710 &  -0.025 \\
22.204 &  2.633 &  -0.077 & 22.112 &  1.713 &  -0.025 \\
21.873 &  2.647 &  -0.079 & 21.837 &  1.716 &  -0.025 \\
21.546 &  2.658 &  -0.080 & 21.552 &  1.718 &  -0.025 \\
\tableline
\end{tabular}
Note: Table 5 is presented in its entirety in the electronic edition
of the Astronomical Journal.  A portion is shown here for guidance
regarding its form and content.  
} \\ \\ \\ \\ \\ \\ \\

\noindent
\parbox{3.25in}{
{\sc Table 6:}NGC~6528 ridge line and reddening coefficients

\begin{tabular}{cccccc}
\tableline
$m_{F606W}$ & $\alpha_{F606W}$ &  & $m_{F814W}$ & $\alpha_{F814W}$ &  \\ 
(mag)       &  (mag)           & $\beta_{F606W}$ & (mag) & (mag) & $\beta_{F814W}$  \\
\tableline
\tableline
24.847 &  2.559 &  -0.066 & 23.908 &  1.680 &  -0.024 \\
24.554 &  2.564 &  -0.066 & 23.710 &  1.686 &  -0.024 \\
24.253 &  2.567 &  -0.067 & 23.508 &  1.690 &  -0.024 \\
23.961 &  2.575 &  -0.068 & 23.309 &  1.695 &  -0.024 \\
23.660 &  2.583 &  -0.069 & 23.104 &  1.700 &  -0.025 \\
23.365 &  2.594 &  -0.071 & 22.900 &  1.704 &  -0.025 \\
23.088 &  2.604 &  -0.072 & 22.696 &  1.707 &  -0.025 \\
22.823 &  2.618 &  -0.074 & 22.493 &  1.710 &  -0.025 \\
22.566 &  2.629 &  -0.076 & 22.288 &  1.712 &  -0.025 \\
22.316 &  2.640 &  -0.078 & 22.088 &  1.714 &  -0.025 \\
\tableline
\end{tabular}

Note: Table 6 is presented in its entirety in the electronic edition
of the Astronomical Journal.  A portion is shown here for guidance
regarding its form and content.  
}

\vskip 0.63in

\noindent
\parbox{3.25in}{
{\sc Table 7:}NGC~6791 ridge line and reddening coefficients

\begin{tabular}{cccccc}
\tableline
$m_{F606W}$ & $\alpha_{F606W}$ &  & $m_{F814W}$ & $\alpha_{F814W}$ &  \\ 
(mag)       &  (mag)           & $\beta_{F606W}$ & (mag) & (mag) & $\beta_{F814W}$  \\
\tableline
\tableline
22.838 &  2.552 &  -0.067 & 22.014 &  1.670 &  -0.023 \\
22.333 &  2.559 &  -0.067 & 21.687 &  1.680 &  -0.024 \\
21.640 &  2.569 &  -0.067 & 21.238 &  1.690 &  -0.024 \\
21.071 &  2.581 &  -0.069 & 20.880 &  1.700 &  -0.025 \\
20.667 &  2.595 &  -0.071 & 20.600 &  1.706 &  -0.025 \\
20.002 &  2.621 &  -0.075 & 20.102 &  1.713 &  -0.025 \\
19.647 &  2.634 &  -0.077 & 19.828 &  1.716 &  -0.025 \\
19.327 &  2.645 &  -0.078 & 19.567 &  1.718 &  -0.025 \\
18.905 &  2.657 &  -0.080 & 19.202 &  1.721 &  -0.025 \\
18.480 &  2.666 &  -0.081 & 18.822 &  1.723 &  -0.025 \\
\tableline
\end{tabular}
Note: Table 7 is presented in its entirety in the electronic edition
of the Astronomical Journal.  A portion is shown here for guidance
regarding its form and content.  
}

\vskip 0.63in

\noindent
\parbox{3.25in}{
{\sc Table 8:}NGC~6341 HB locus and reddening coefficients

\begin{tabular}{cccccc}
\tableline
$m_{F606W}$ & $\alpha_{F606W}$ &  & $m_{F814W}$ & $\alpha_{F814W}$ &  \\ 
(mag)       &  (mag)           & $\beta_{F606W}$ & (mag) & (mag) & $\beta_{F814W}$  \\
\tableline
\tableline
16.858 &  2.801 &  -0.092 & 18.074 &  1.752 &  -0.025 \\
16.838 &  2.801 &  -0.092 & 17.993 &  1.752 &  -0.025 \\
16.245 &  2.801 &  -0.092 & 17.397 &  1.752 &  -0.025 \\
16.769 &  2.801 &  -0.092 & 17.905 &  1.752 &  -0.025 \\
16.789 &  2.801 &  -0.092 & 17.916 &  1.752 &  -0.025 \\
16.943 &  2.801 &  -0.092 & 18.068 &  1.752 &  -0.025 \\
16.791 &  2.801 &  -0.092 & 17.916 &  1.752 &  -0.025 \\
16.739 &  2.801 &  -0.092 & 17.862 &  1.752 &  -0.025 \\
16.722 &  2.799 &  -0.092 & 17.841 &  1.751 &  -0.025 \\
16.586 &  2.799 &  -0.092 & 17.703 &  1.751 &  -0.025 \\
\tableline
\end{tabular}
Note: Table 8 is presented in its entirety in the electronic edition
of the Astronomical Journal.  A portion is shown here for guidance
regarding its form and content.  
} \\ \\ \\

\noindent
\parbox{3.25in}{
{\sc Table 9:}NGC~6752 HB locus and reddening coefficients

\begin{tabular}{cccccc}
\tableline
$m_{F606W}$ & $\alpha_{F606W}$ &  & $m_{F814W}$ & $\alpha_{F814W}$ &  \\ 
(mag)       &  (mag)           & $\beta_{F606W}$ & (mag) & (mag) & $\beta_{F814W}$  \\
\tableline
\tableline
17.674 &  2.802 &  -0.092 & 18.954 &  1.752 &  -0.025 \\
17.740 &  2.802 &  -0.092 & 18.979 &  1.752 &  -0.025 \\
17.271 &  2.802 &  -0.092 & 18.507 &  1.752 &  -0.025 \\
18.010 &  2.802 &  -0.092 & 19.244 &  1.752 &  -0.025 \\
17.569 &  2.802 &  -0.092 & 18.797 &  1.752 &  -0.025 \\
17.696 &  2.802 &  -0.092 & 18.923 &  1.752 &  -0.025 \\
17.744 &  2.802 &  -0.092 & 18.967 &  1.752 &  -0.025 \\
17.315 &  2.802 &  -0.092 & 18.536 &  1.752 &  -0.025 \\
16.533 &  2.802 &  -0.092 & 17.741 &  1.752 &  -0.025 \\
17.530 &  2.802 &  -0.092 & 18.737 &  1.752 &  -0.025 \\
\tableline
\end{tabular}
Note: Table 9 is presented in its entirety in the electronic edition
of the Astronomical Journal.  A portion is shown here for guidance
regarding its form and content.  
}

\vskip 0.63in

\noindent
\parbox{3.25in}{
{\sc Table 10:}NGC~104 HB locus and reddening coefficients

\begin{tabular}{cccccc}
\tableline
$m_{F606W}$ & $\alpha_{F606W}$ &  & $m_{F814W}$ & $\alpha_{F814W}$ &  \\ 
(mag)       &  (mag)           & $\beta_{F606W}$ & (mag) & (mag) & $\beta_{F814W}$  \\
\tableline
\tableline
14.075 &  2.678 &  -0.083 & 14.484 &  1.722 &  -0.025 \\
13.874 &  2.678 &  -0.083 & 14.279 &  1.722 &  -0.025 \\
14.028 &  2.678 &  -0.083 & 14.432 &  1.722 &  -0.025 \\
14.056 &  2.678 &  -0.083 & 14.460 &  1.722 &  -0.025 \\
13.998 &  2.678 &  -0.083 & 14.400 &  1.722 &  -0.025 \\
14.028 &  2.678 &  -0.083 & 14.430 &  1.722 &  -0.025 \\
14.085 &  2.678 &  -0.083 & 14.485 &  1.722 &  -0.025 \\
13.940 &  2.678 &  -0.083 & 14.339 &  1.722 &  -0.025 \\
14.068 &  2.678 &  -0.083 & 14.466 &  1.722 &  -0.025 \\
14.074 &  2.678 &  -0.083 & 14.472 &  1.722 &  -0.025 \\
\tableline
\end{tabular}

Note: Table 10 is presented in its entirety in the electronic edition
of the Astronomical Journal.  A portion is shown here for guidance
regarding its form and content.  
}

\vskip 0.63in

\noindent
\parbox{3.25in}{
{\sc Table 11:}NGC~5927 HB locus and reddening coefficients

\begin{tabular}{cccccc}
\tableline
$m_{F606W}$ & $\alpha_{F606W}$ &  & $m_{F814W}$ & $\alpha_{F814W}$ &  \\ 
(mag)       &  (mag)           & $\beta_{F606W}$ & (mag) & (mag) & $\beta_{F814W}$  \\
\tableline
\tableline
16.407 &  2.671 &  -0.082 & 16.402 &  1.721 &  -0.025 \\
16.430 &  2.671 &  -0.082 & 16.421 &  1.721 &  -0.025 \\
16.370 &  2.671 &  -0.082 & 16.361 &  1.721 &  -0.025 \\
16.506 &  2.671 &  -0.082 & 16.490 &  1.721 &  -0.025 \\
16.403 &  2.671 &  -0.082 & 16.384 &  1.721 &  -0.025 \\
16.534 &  2.671 &  -0.082 & 16.515 &  1.721 &  -0.025 \\
16.407 &  2.671 &  -0.082 & 16.386 &  1.721 &  -0.025 \\
16.503 &  2.671 &  -0.082 & 16.479 &  1.721 &  -0.025 \\
16.452 &  2.671 &  -0.082 & 16.428 &  1.721 &  -0.025 \\
16.520 &  2.671 &  -0.082 & 16.495 &  1.721 &  -0.025 \\
\tableline
\end{tabular}
Note: Table 11 is presented in its entirety in the electronic edition
of the Astronomical Journal.  A portion is shown here for guidance
regarding its form and content.  
} \\ \\ \\ \\ \\ \\

\noindent
\parbox{3.25in}{
{\sc Table 12:}NGC~6528 HB locus and reddening coefficients

\begin{tabular}{cccccc}
\tableline
$m_{F606W}$ & $\alpha_{F606W}$ &  & $m_{F814W}$ & $\alpha_{F814W}$ &  \\ 
(mag)       &  (mag)           & $\beta_{F606W}$ & (mag) & (mag) & $\beta_{F814W}$  \\
\tableline
\tableline
16.784 &  2.669 &  -0.081 & 16.654 &  1.720 &  -0.025 \\
16.766 &  2.669 &  -0.081 & 16.636 &  1.720 &  -0.025 \\
16.728 &  2.669 &  -0.081 & 16.597 &  1.720 &  -0.025 \\
16.954 &  2.669 &  -0.081 & 16.818 &  1.720 &  -0.025 \\
16.981 &  2.669 &  -0.081 & 16.841 &  1.720 &  -0.025 \\
16.813 &  2.669 &  -0.081 & 16.670 &  1.720 &  -0.025 \\
16.835 &  2.669 &  -0.081 & 16.691 &  1.720 &  -0.025 \\
16.927 &  2.669 &  -0.081 & 16.778 &  1.720 &  -0.025 \\
16.876 &  2.669 &  -0.081 & 16.727 &  1.720 &  -0.025 \\
16.935 &  2.669 &  -0.081 & 16.784 &  1.720 &  -0.025 \\
\tableline
\end{tabular}
Note: Table 12 is presented in its entirety in the electronic edition
of the Astronomical Journal.  A portion is shown here for guidance
regarding its form and content.  
}

\vskip 0.63in

\noindent
\parbox{3.25in}{
{\sc Table 13:}NGC~6791 HB locus and reddening coefficients

\begin{tabular}{cccccc}
\tableline
$m_{F606W}$ & $\alpha_{F606W}$ &  & $m_{F814W}$ & $\alpha_{F814W}$ &  \\ 
(mag)       &  (mag)           & $\beta_{F606W}$ & (mag) & (mag) & $\beta_{F814W}$  \\
\tableline
\tableline
14.455 &  2.623 &  -0.074 & 14.533 &  1.719 &  -0.025 \\
14.528 &  2.620 &  -0.073 & 14.597 &  1.719 &  -0.025 \\
14.524 &  2.620 &  -0.073 & 14.583 &  1.719 &  -0.025 \\
14.480 &  2.615 &  -0.073 & 14.502 &  1.717 &  -0.025 \\
14.537 &  2.612 &  -0.073 & 14.501 &  1.713 &  -0.025 \\
\tableline
\end{tabular}
Note: Table 13 is presented in its entirety here, but for completeness
a machine-readable table is available in the electronic edition of the
Astronomical Journal, matching the format of Tables 2--12.  
}

\vskip 0.3in

In Figure 7, we show all of the ridge lines and HB loci transformed to
a distance of 10 pc and $E(B-V)=0$~mag, using the information in
Tables 1--13.  Specifically, for each cluster, we calculated
$M_{F606W} \equiv m_{F606W} - A_{F606W} - (m-M)_0$ and $M_{F814W}
\equiv m_{F814W} - A_{F8146W} - (m-M)_0$, with $A = \alpha \times
E(B-V) + \beta \times [E(B-V)]^2$ and $(m-M)_0 = (m-M)_V - 3.1 \times E(B-V)$.
When placed in the same reference frame, the clusters fall at their
expected relative locations, given their relative ages and
metallicities (Table 1).  Although the distances for these clusters
were determined via a variety of methods, there is clear consistency
in the relative distances and reddenings, given the excellent
agreement at the HB.  Note that the NGC~6528 and NGC~6791 photometry
reaches only $\sim 3$~mag below the turnoff, and so their fiducials
are truncated above the point where the main sequence begins to
steepen (in contrast with the other clusters, which have fiducials reaching
$\sim 5$~mag below the turnoff); this variation in depth exaggerates
the impression that the fiducials are diverging at the faint limit.

\section{Summary}

Using observations of Galactic clusters spanning a wide range in metallicity,
we provide ridge lines and HB loci in the two most popular ACS/WFC
bandpasses, along with coefficients for transforming these fiducials
to a reference frame with an arbitrary reddening and distance.  We also
provide the algorithm for transforming the VRI to the
ACS bandpasses, which provides good agreement ($\lesssim 0.02$~mag) with
the cluster data from a point 1.5~mag below the main sequence turnoff to
a point near the tip of the RGB.  The empirical fiducials and the 
isochrone transformations should be useful in the analysis of stellar
population data obtained using ACS.

\acknowledgements

Support for proposals 9453 and 10265 was provided by NASA through a
grant from STScI, which is operated by AURA, Inc., under NASA contract
NAS 5-26555.  We are grateful to P.\ Stetson for providing his codes
and assistance.  We are also grateful to F.\ Castelli for making her
grids of synthetic spectra available to us.  We thank the members of
the scheduling and operations teams at STScI (especially P.\ Royle,
D.\ Taylor, and D.\ Soderblom) for their efforts in executing our
large {\it HST} programs.

\vskip 0.2in

\epsfxsize=6.5in \epsfbox{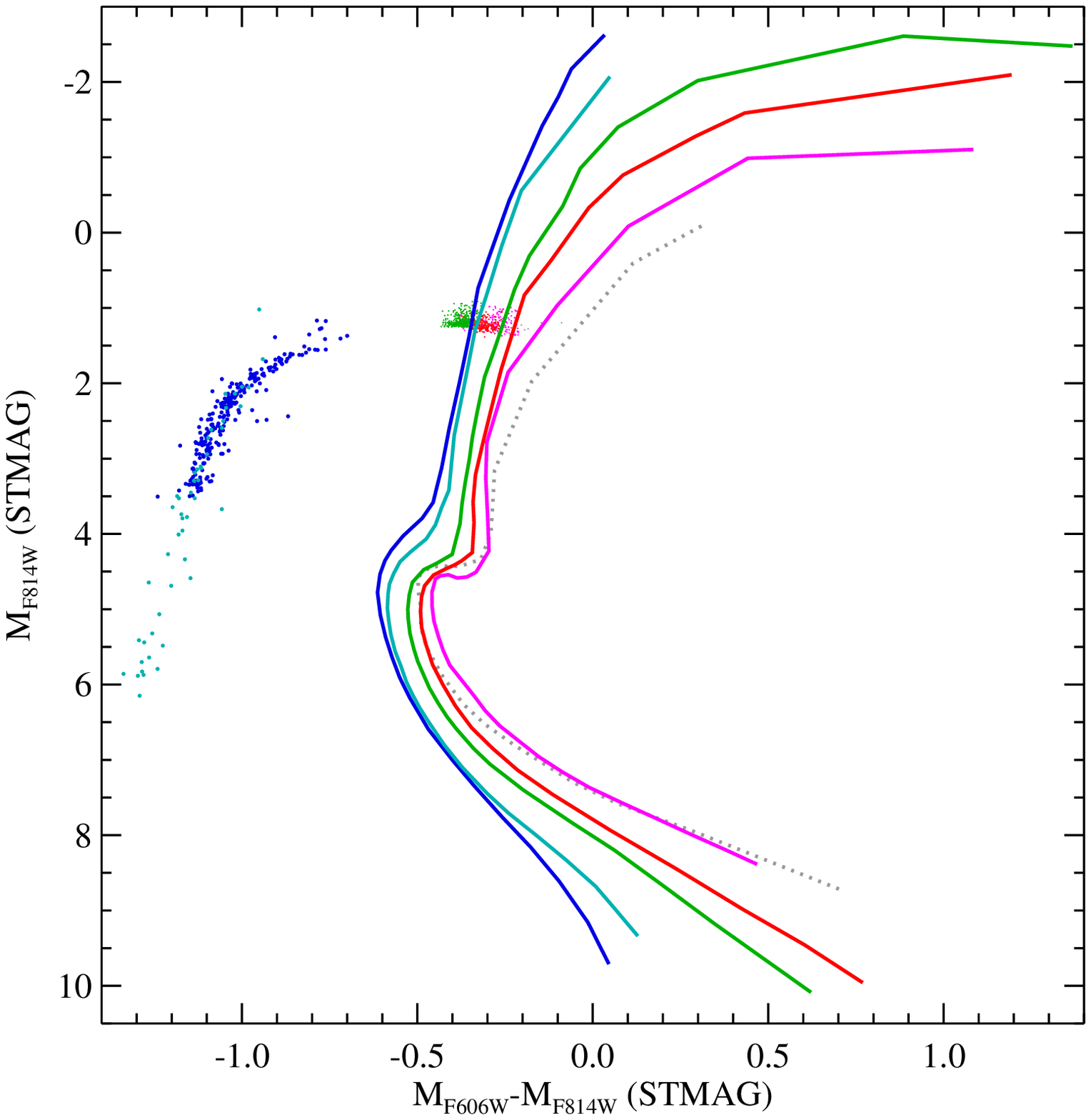}

\parbox{6.5in}{\small {\sc Fig.~7--} 
The ridge lines and HB loci of Figures 1--6, shifted to a
distance of 10 pc and no reddening, using the reddening
coefficients of Tables 2--13 and the distance and reddening
specified in Table 1.  The globular cluster fiducials
({\it color points and solid curves}) and open cluster ({\it grey
points and dashed curve}) fall where one would expect relative to each
other, given their relative ages and metallicities (Table 1).  The HB
loci also show good agreement, which suggests the relative distance
moduli and extinctions are consistent, even though they were derived
via a variety of techniques in the literature (fitting subgiants,
white dwarfs, and HB stars).}

\end{document}